\documentclass{emulateapj}

\def\Msun{{\hbox {M$_\odot$}}} 
\newcommand{\HII}{H\,{\sc ii}}
\newcommand{\OIII}{O\,{\sc iii}}
\newcommand{\CII}{C\,{\sc ii}}
\newcommand{\NII}{N\,{\sc ii}}
\newcommand{\OI}{O\,{\sc i}}
\newcommand{\NIII}{N\,{\sc iii}}

\slugcomment{Pre-print version, accepted in ApJ main journal, 2006 January 12}

\shorttitle{Water Vapor around Sgr B2}
\shortauthors{Cernicharo, Goicoechea, Pardo \& Asensio Ramos}

\begin{document}

\title{Warm Water Vapor around Sagittarius B2\altaffilmark{1}}

\author{Jos\'e Cernicharo, Javier R. Goicoechea\altaffilmark{2}, 
Juan R. Pardo} 

\affil{Departamento de Astrof\'isica Molecular e Infrarroja, IEM,
CSIC, Serrano 121, 28006, Madrid, Spain}

\and

\author{Andr\'es Asensio-Ramos}
\affil{Instituto de Astrof\'{\i}sica de Canarias, E-38205, La Laguna,
       Tenerife, Spain }

\email{cerni@damir.iem.csic.es}

\altaffiltext{1}{Based on observations with ISO, 
an ESA project with instruments funded by ESA Member States 
(especially the PI countries: France, Germany, the Netherlands 
and the United Kingdom) and with participation of ISAS and NASA.} 

\altaffiltext{2}{Present address: Laboratoire d'\'{E}tude du Rayonnement et de la 
Mati\`ere, UMR 811, CNRS,  Observatoire de
Paris et \'{E}cole Normale Sup\'{e}rieure, 24 rue Lhomond,  75231 Paris Cedex 05, France}

\begin{abstract}

In the region of Sgr B2 there are several condensations heated externally 
by nearby hot stars. Therefore H$_2$O far--IR lines are expected to probe only 
an external low--density and high temperature section of these
condensations, 
whereas 
millimeter-wave lines can penetrate deeper into them where the density is higher 
and T$_k$ lower. We have conducted a study combining H$_2$O lines in both 
spectral regions. First, $\textit{Infrared Space Observatory}$ observations 
of several H$_2$O thermal lines seen in absorption
toward Sgr~B2(M) at a spectral resolution of
$\sim$35~km~s$^{-1}$ have been analyzed. Second, 
an \textit{IRAM}--30m telescope map of the 
para--H$_2$O $3_{13}-2_{20}$ line at 183.31~GHz, seen in emission, has also been 
obtained and analyzed. 
The H$_2$O lines seen in absorption are optically thick and are formed in
the outermost gas of the condensations  
in front of the far--IR continuum sources. They probe a maximum 
visual extinction of  $\sim$5 to 10 mag. Radiative transfer models indicate that
these lines are quite insensitive to temperature and gas density, and
that IR photons from the dust play a dominant
role in the excitation of the involved H$_2$O rotational levels. In order to get 
the physical conditions of the absorbing gas
we have also analyzed the CO emission toward Sgr~B2(M). We conclude, based 
on the observed CO $J$=7--6 line at
806.65~GHz with the \textit{Caltech~Submillimeter~Observatory}, and the lack
of emission from the far--IR CO lines, that the gas density has to be
lower than $\sim$10$^4$~cm$^{-3}$. Using the values obtained for the kinetic 
temperature and gas density
from OH, CO, and other molecular species, we derive a water column
density of  (9$\pm$3)$\times$10$^{16}$~cm$^{-2}$ in the absorbing gas.
Hence, the water vapor abundance in this region, $\chi$(H$_2$O), is
$\simeq$(1-2)$\times$10$^{-5}$. The relatively low H$_2$O/OH abundance ratio 
in the region, $\simeq$2-4, is a signature of UV photon dominated surface
layers traced by far--IR observations. As a consequence
the temperature of the absorbing gas is high, T$_K\simeq$300-500 K, which 
allows very efficient neutral--neutral reactions producing H$_2$O and OH.
On the other hand, the 183.31 GHz data provide a much better spatial and spectral 
resolution than the far-IR ISO data. This maser line allows to trace water deeper
into the cloud, i.e., the inner, denser ($n(H_2)$$\ge$10$^{5-6}$~cm$^{-3}$) 
and colder (T$_k$$\sim$40 K) gas. The emission is very strong toward the cores. 
The derived water vapor abundance for this component is a few$\times$10$^{-7}$. There  
is also moderate extended emission around Sgr~B2 main condensations, a fact that supports 
the water vapor abundance derived from far--IR H$_2$O lines for the outer gas.

\end{abstract}


\section{Introduction}
\label{sct:intro}
The determination of water  abundance in space is a long standing 
problem in Astronomy. Theoretical models predict that  water
can be the most abundant species in 
\textit{warm} molecular clouds after H$_2$ and CO \citep{Neu95}.
Therefore, the determination of its spatial distribution and
abundance contributes to a better knowledge
of the chemical  and physical processes that take place
in the interstellar medium (ISM).

Unfortunately, water is an abundant molecule in our  atmosphere making particularly 
difficult the observation  of its rotational lines and vibrational bands
from Earth. Even so, some 
observations of water lines have been performed from ground--based
and airborne telescopes: the $6_{16}-5_{23}$  at
22~GHz \citep{Che69}, the $5_{15}-4_{22}$  at
325~GHz \citep{Men90}, the $10_{29}-9_{36}$ 
at 321~GHz \citep{Men90b} and the $3_{13}-2_{20}$ 
at 183.31~GHz (Watson et al. 1980; Cernicharo et al. 1990, 1994,
1996, 1999a; Gonz\'alez-Alfonso et al., 1998).
Due to the maser nature of these lines, the analysis and interpretation
of the spectra is not obvious.
Among these lines, only two have been used to map the extended emission of water vapor in Orion: 
the $3_{13}-2_{20}$ -hereafter the 183 GHz line- \citep{Cer94} 
and the $5_{15}-4_{22}$ at 325 GHz \citep{Cer99a}. 
However, although we know from ISO and SWAS observations 
that water is extended ($\sim$25$'$$\times$25$'$) in Sgr~B2  
(Cernicharo et al. 1997, Neufeld et al. 2003, Goicoechea et al. 2004), little 
is known about its excitation conditions  
and its detailed spatial distribution. An alternative to indirectly estimate 
the water abundance in the Galactic Center (GC) is to use related species 
such as HDO \citep{Jac90,Com03} or H$_3$O$^+$ \citep{Phi92,Goi01}.
In none of these cases, the determination of $\chi$(H$_2$O) is straightforward.

The ISO mission \citep{Kes96},
and specially, the \textit{Long Wavelength Spectrometer}, LWS, \citep{Cle96},
and the \textit{Short Wavelength Spectrometer}, SWS, \citep{Thi96},
have provided a unique opportunity to observe 
several H$_2$O lines in a great variety of astronomical environments. 
Nevertheless, the majority 
of these observations were performed at the low spectral resolution
of the grating mode  ($\sim$1000~km~s$^{-1}$), which produces a critically strong 
dilution in the search for molecular features in most  ISM sources.
Nevertheless, the Sgr~B2 cloud has been analyzed
and studied in detail with the LWS/\textit{Fabry--Perot}, which provided
a velocity resolution of $\sim$35~km~s$^{-1}$
(Goicoechea et al. 2004, hereafter G04).

Opposite to what is found toward other star forming regions such as Orion
\citep{Cer99b}, the observations of the $2_{12}-1_{01}$ line  at
$\sim$179.5~$\mu$m in Sgr B2 show that the line 
appears in absorption rather than in emission \citep{Cer97}.
Afterward, the launch of the \textit{Submillimeter Wave Astronomy
Satellite}, SWAS, \citep{Mel00}, and ODIN \citep{Nor03}  allowed the
observation of the $1_{10}-1_{01}$ fundamental transition of both
H$_{2}^{16}$O at 557 
and H$_{2}^{18}$O at 548~GHz, first detected by the \textit{Kuiper Airborne
Observatory}, KAO, \citep{Zmu95}.
Although the velocity resolution is $\sim$1~km~s$^{-1}$, the large
beam of SWAS ($\sim$4$'$) makes these observations more sensitive
to the cold and less dense gas.
  SWAS observations have provided a reliable estimate of the
water vapor abundance in the low excitation clouds located in the line of sight
toward Sgr~B2 \citep{Neu00,Neu03}.  However, the fact that only the
ground-state absorption line is detected makes difficult a detailed study of 
water vapor excitation mechanisms in Sgr~B2 itself. This is 
the most massive cloud in the Galaxy, with $\sim$10$^7$ \Msun,
\citep{Lis90}, and a paradigmatic object in the GC  region
as its geometrical properties, physical conditions and chemical
characteristics make it a miniature galactic nucleus with 
$\sim$15$'$ extent  (G04 and references therein).

The main star-forming regions in Sgr~B2 are located within
three dust condensations, labeled as (N), (M) and (S).
These condensations are embedded in a $\sim$10~pc moderate--density  
($n(H_2)=10^5-10^6$~cm$^{-3}$) cloud \citep{Lis91,Hut93}. 
In addition, these structures are surrounded
by lower density components
of warm (T$_k$$\geq$100 K) gas, hereafter the Sgr~B2 warm envelope.
These conditions have been derived mainly from absorption observations
of NH$_3$ metastable lines \citep{Wil82,Hut95}, OH lines 
(Goicoechea \& Cernicharo, 2001; 2002), and H$_2$CO lines \citep{Mar90}
in the radio domain. Nevertheless, the warm and low density gas is
poorly traced by radio observations of other molecular species 
(generally excited by collisions in the denser regions and thus observed in emission).
However, the warm envelope represents the strongest contribution to the absorption
features produced by many light hydrides in the far-IR spectrum
of Sgr~B2 \citep{G04}. 
The origins of the observed rich chemistry  and the  heating 
mechanisms in the Sgr~B2 warm envelope are a subject of intense debate.
The matter is complicated  due to the different observational signatures to be
integrated in the same picture: high temperatures derived from NH$_3$
absorption lines  \citep{Hut95,Cec02}, fine
structure emission from the photo--ionized and photo--dissociated gas \citep{G04}, SiO
and X--ray distribution \citep{Mar00}, etc. 

In all possible scenarios, water plays a significant role. 
Several mechanisms allow its formation and survival in the warm envelope.
Dissociative recombination of H$_3$O$^+$ leads to the production
of H$_2$O and OH. This processes depend on the specific $f_{H_2O}$ branching ratio
for the H$_2$O formation channel. Unfortunately, the determination of $f_{H_2O}$
with different experimental procedures has also yielded different values, 
from $f_{H_2O}$=0.05 \citep{Wil96} to $f_{H_2O}$=0.25 
\citep{Jen00}, while most chemical models have used $f_{H_2O}$$\sim$0.35.
In addition, water could also be produced in the gas phase 
by the endothermic reaction: 
\begin{equation}
OH + H_2 \rightarrow H_2O + H 
\end{equation}
However, the gas temperature must exceed $\sim$300~K to overcome the
activation barrier \citep{Neu95}.
At these temperatures, the reaction:
\begin{equation}
O + H_2  \rightarrow OH + H
\end{equation}
also contributes to the formation of OH. Therefore, H$_2$O and OH column
densities can be used to determine the role of the neutral-neutral reactions
in their formation/destruction routes. 
Still, the exact H$_2$O/OH ratio will be determined by $f_{H_2O}$, the temperature
and by photodissociation processes if UV radiation is present. 
As an example, Neufeld et al. \citep{Neu03} studied a diffuse cloud 
($G_0$$\sim$1, $n(H)$=100~cm$^{-3}$) toward W51 and showed that the presence
of  a warm gas component (T$_k$$\gtrsim$400~K) could explain the observed variations
of the H$_2$O/OH ratio respect to other diffuse clouds.

Finally, high oxygen depletion onto water ice
mantles in dust grains could have taken place during the evolution
of the cool gas in Sgr~B2. 
Photodesorption and/or evaporation for dust temperatures above $\sim$90~K,
could release some water back into the gas phase enhancing 
the H$_2$O  abundance expected from pure gas--phase formation \citep{Ber00}.
However, gas and dust are thermally decoupled in the outer layers
of Sgr~B2, where the dust temperatures are significantly  
lower, T$_d$$\simeq$20--30~K \citep{Gor93,G04},
than gas temperatures, T$_k$$\simeq$300~K \citep{Goi02, Cec02}.
Thus, T$_d$ seems too low to produce significant
evaporation of water ice mantles.
Therefore, a detailed  study  of the far--IR H$_2$O  lines and of the 
183.31~GHz extended emission is needed to constrain the water abundance and 
the physical characteristics of the absorbing/emitting region.

In this work we present and analyze the far-IR observations of
several thermal lines of water vapor  toward Sgr~B2(M) and the first
map of the 183.31~GHz maser emission of para--H$_2$O around Sgr~B2
main condensations.
The layout of the paper is as follows:
In Section \ref{sct:observ} we summarize the far--IR, submm and mm observations and  
data reduction. The spectra and maps are presented in 
Sec.\ref{sct:results}. Section \ref{sct:analysis} is 
devoted to the analysis of CO observations (\ref{ssct:co}) and 
water vapor observations (\ref{ssct:h2o}) with different radiative 
transfer methods. The main implications of our work are discussed in Sec.
\ref{sct:discussion}, where 
photochemistry models for H$_2$O and OH are also presented.
A summary is given in Sec. \ref{sct:summary}.

\section{Observations and Data Reduction}
\label{sct:observ}
\subsection{Far-IR observations}
\label{ssct:farir}
Most pure rotational lines of water vapor that play a role in the
radiative heating and cooling of the dense ISM appear in the 
Terahertz domain. Thus, airborne or satellite observations are needed
to avoid the Earth's atmosphere blocking.
In particular, many pure rotational lines of H$_2$O appear in the far--IR
coverage of the LWS spectrometer \citep{Cle96} on board ISO \citep{Kes96}.
We have used the LWS  Fabry-Perot (FP) instrument to search for   
H$_2$O and H$_2^{18}$O lines toward Sgr~B2(M).
The LWS/FP spectral resolution is $\lambda/\Delta\lambda\simeq$
7000-1000 and has a circular aperture of about  
80$''$ in diameter. The majority of detected lines
have been observed in the time awarded to our ISO proposals. 
However, an  extensive inspection of the public ISO data 
base\altaffilmark{3} has been carried out in order
to examine and average all available water lines. 
Most of the  water lines were present in
the Astronomical Observation Template (AOT)~LWS04 observations, which give a
large spectral sampling, wavelength precision and S/N ratios.
The number of scans in this mode was $\geq$12 depending on
the expected absorption produced by the different species.

\footnotetext[3]{The ISO data base observations (TDTs) presented in these work are
32201428,
32201429,
46201118,
46201123,
46900332,
47600907,
47600908,
47600909,
47601001 and
49800301.}

These LWS products have been processed and compared through 
Off-Line-Processing (OLP) pipeline 
6.0 and 10.1 versions. There are no major differences except
that recent pipelines produce $<$10\% less absorption in some lines
due to continuum level differences from one OLP to another.
The data were analyzed using the ISO spectrometers data reduction package 
ISAP\altaffilmark{4}.
The mean  FP continuum flux of each line deviates by $<$20\% and
this can be taken as the flux calibration error \citep{Swi98,Gry03}.
After checking the continuum level of the AOT LWS01 observations, a
polynomial baseline was fitted to each spectra and adopted as
FP continuum level. 

\footnotetext[4]{ISAP is a joint 
development by the LWS and SWS Instrument Teams and Data Centers.
Contributing institutes are CESR, IAS, IPAC, MPE, RAL and SRON.}

\begin{figure*} [t] 
\centering

\includegraphics[angle=0,width=14cm]{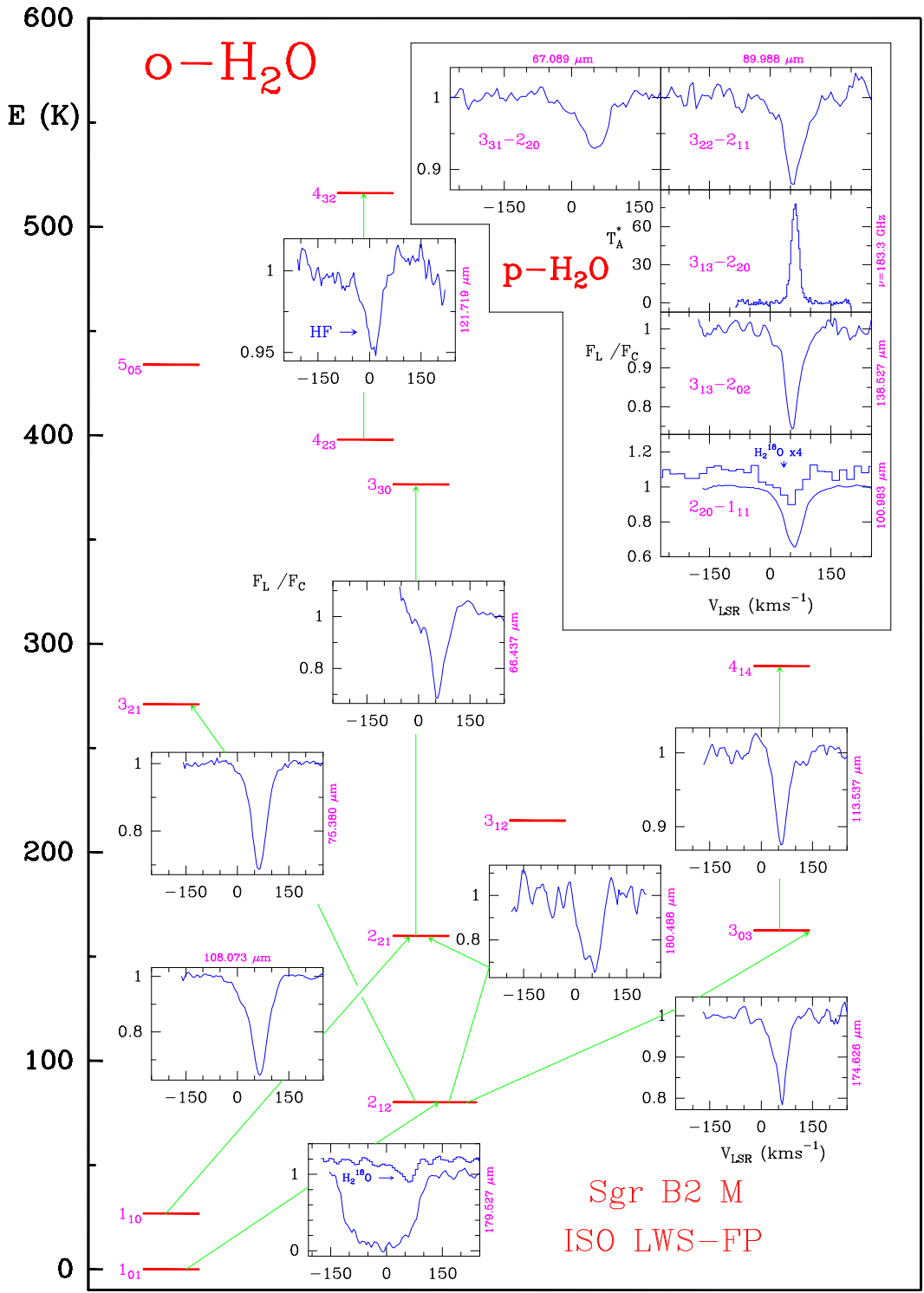}
\caption{ISO/LWS/FP observations of H$_{2}^{16}$O and H$_{2}^{18}$O
toward Sgr~B2(M) and rotational energy diagram of ortho--H$_{2}^{16}$O.
The ordinate scale corresponds to F/F$_c$ and the
abscissa to the wavelength in microns. All the far-IR water lines appear in absorption.
The emission line corresponds to the 3$_{13}$--2$_{20}$ maser transition
of para--H$_{2}^{16}$O at 183.31~GHz (observed with the IRAM--30m telescope
at Pico Veleta, Spain).}
\label{obs_iso_agua}
\end{figure*}

\begin{figure*}[]
\centering
\includegraphics[angle=-90, width=16.5cm]{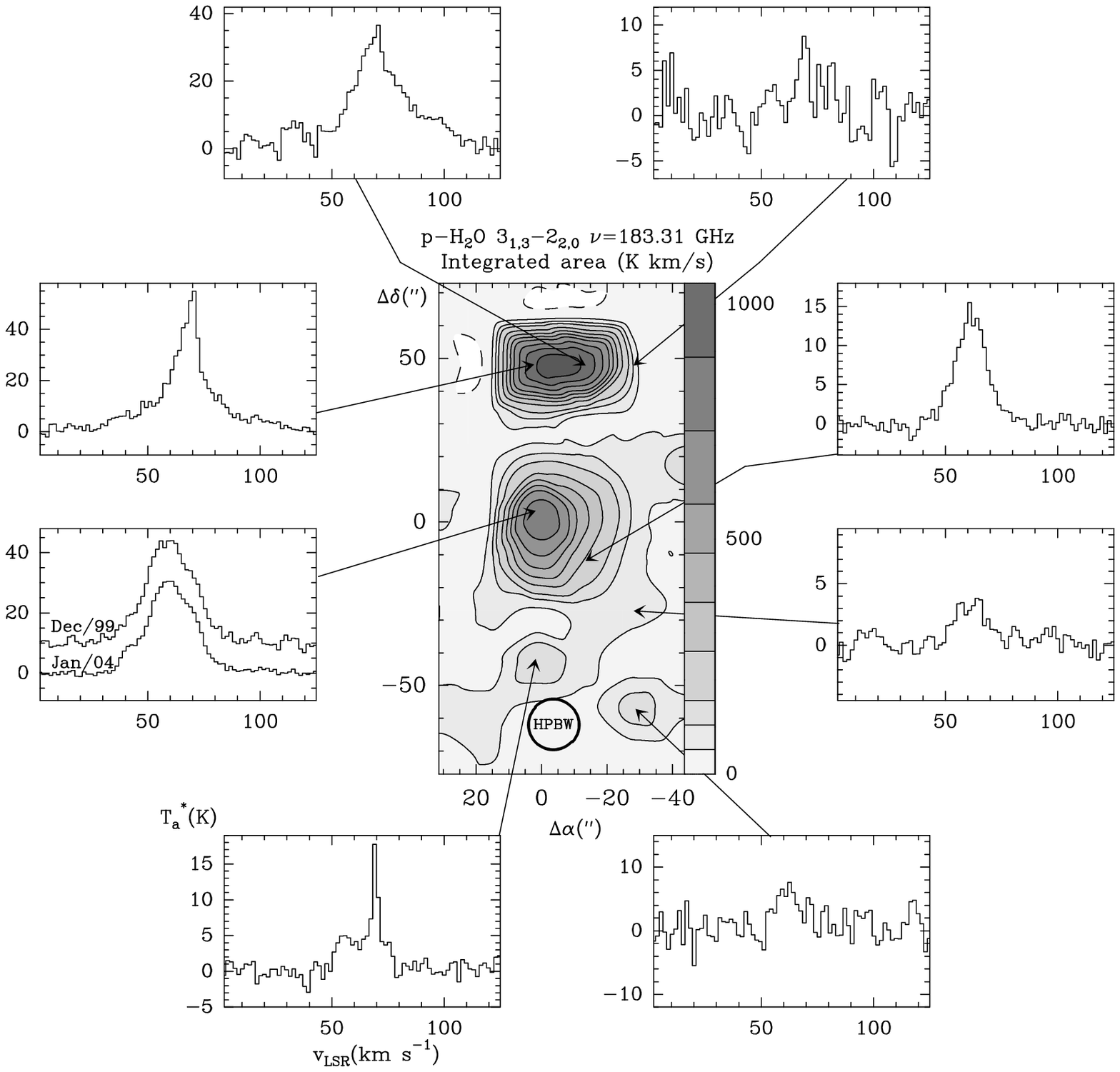}
\caption{Map of the 3$_{13}$--2$_{20}$ line
of para--H$_{2}^{16}$O at 183.31~GHz around Sgr~B2 main condensations. Contours in
K~km~s$^{-1}$ are indicated in the figure. Different line profiles at 
significant positions of the map are also shown. The intensity scale is
in T$^{*}_{A}$ and the abscissa is the LSR velocity in km~s$^{-1}$. The map
is centered at the position of Sgr~B2(M). The spectra were obtained in two different 
runs in Dec. 1999 and Jan. 2004. The fluxes were scaled using the (0'',5'') position  
as reference. Note that the line profiles in that reference position did not change 
in more than 4 years.}
\label{mapa_183}
\end{figure*}

\subsection{Millimeter and submillimeter observations}
\label{ssct:mmsubmm}
The water line at 183.31~GHz can be observed from high mountain-top sites 
under very dry conditions \citep{Cer90}. The observations in 
the direction of Sgr~B2 were achieved with the 
IRAM-30m telescope, with a half--power--beam--width (HPBW) of
$\sim$13$''$, in December 1999 and January 2004.
The source transits at 24 degrees elevation at Pico Veleta, but even so the
average atmospheric opacity over a bandwidth of 0.5 
GHz centered at 183.3~GHz was 1.7, i.e. an atmospheric water column of 0.9~mm 
along the line of sight. We used and SIS receiver designed to cover the band 130-184~GHz. In
its higher end, the receiver temperature is about 160~K and the sideband 
rejection is at least 18~dB. The backend used was a 512 two-pole filter with 
half-power widths and spacings equal to 1.0 MHz. With these conditions, and in 
this configuration, system temperatures were typically around 3000~K. The 
pointing was checked using the already known strong emission of W49N \citep{Cer90}.
Once it was verified that the Sgr~B2 emission was 
quite compact, the map was carried out in wobbler-switching mode 
in order to obtain very flat baselines. 
Observations were also tried  in January 2001, 2002 and 2004. In the 2001 run 
the weather was poor while in the last run
it was possible to have a 30 minutes window of observing time with good
atmospheric transmission. These  runs allowed us to check that the line
profile in the position peak of Sgr~B2(M) was the same of December 1999
and to confirm that the extent of the water emission toward this condensation was
larger 
than the beam size. In the January 2004 observing run we also had very 
good atmospheric transmission
to perform observations around Sgr~B2(N). However, the measured line fluxes were 
lower than in previous observations by a constant factor. As the line profiles 
did not change, we scaled the 1999 observations to those the reference taken in the 
(0'',+5'') position in 2004 (see Fig.~\ref{mapa_183}).

The CO 7-6 (806~GHz) observations were performed on March 10, 2002
with the 10.4~m \textit{Caltech Submillimeter Observatory} (CSO) located
at the summit of Mauna Kea (Hawaii). The receiver is a
helium-cooled SIS mixer operating in double-sideband mode (DSB) providing
an instantaneous bandwidth of 0.95 GHz and
designed to fully cover the 780-910~GHz atmospheric window (see Kooi et 
al., 2000). The pointing was checked using a strong CO point-like source in the 
nearby W28 molecular cloud, and was kept within 3$''$ accuracy for a 
HPBW of $\sim$10$''$. 
Two different acousto-optic spectrometer (AOS) backends were used with 2048 and
1024 channels respectively for a total spectral coverage
of 1.5 and 0.5~GHz in each case. The zenith atmospheric water vapor column
was $\sim$0.5~mm during the observations, resulting on system temperatures
ranging from 6000 to 10000~K depending on the receiver and the elevation. CO
emission is very extended around Sgr~B2, so we had to perform position
switched
scans setting the off position 1 degree away in azimuth. 
An additional 5\% increase had to be applied to T$_{A}^{*}$ in order
to second-order correct the standard chopper-wheel'' calibration
method at high frequencies with large atmospheric opacities (Pardo et al. 2005).

\section{Results}
\label{sct:results}
The far--IR spectrum of Sgr~B2(M) is dominated by the absorption produced by
NH$_3$, OH and H$_2$O rotational lines \citep{G04}. The detected far-IR lines
of water vapor are shown in Fig.~\ref{obs_iso_agua}.
H$_2$O spectroscopical and observational data are tabulated 
in Table~\ref{tab-water_lines}.
Except the  $2_{12}-1_{01}$
ground state line of $o$-H$_2$O at $\sim$179.5~$\mu$m ($\sim$1669.9~GHz), all water 
lines have a similar profile  and are centered  at Sgr~B2(M) velocities. 
The $\sim$179.5~$\mu$m line is
saturated and absorbs from --150 to +100~km~s$^{-1}$, which therefore includes the
water vapor located in the foreground gas toward the GC
and the warm gas around Sgr~B2(M). 
The widespread absorption produced by the $2_{12}-1_{01}$ line has been 
previously presented 
in Cernicharo et al. (1997) and enlarged in G04 (9$'$$\times$27$'$), 
while the $1_{10}-1_{01}$  absorption  has been mapped (26$'$$\times$19$'$) 
by SWAS (Neufeld et al. 2003).
These observations probe that low excitation  H$_2$O is present
in the clouds intersecting the line of sight toward large areas of Sgr~B2.

The average velocity of all  H$_2$O lines observed with
the ISO/LWS--FP is $+$60$\pm$5~km~s$^{-1}$, in agreement with the 
velocity of other related oxygen  species
such as H$_3$O$^+$ or OH
(Goicoechea \& Cernicharo, 2001; 2002).  
Taking into account the wavelength calibration error of the LWS/FP
instrument, 
this velocity is compatible with the expected $+$65~km~s$^{-1}$ cloud seen in radio observations
\citep{Hut95}. We note, however, that velocities 
close to $+$60~km~s$^{-1}$ are also associated with gas surrounding most 
Sgr~B2 continuum sources, so that the bulk of the H$_2$O absorption
can arise from them. 

Possible overlapping with other molecular species  occurs at some  wavelengths.
In particular, the o--H$_2$O 4$_{32}$--4$_{23}$ line is blended with 
HF $J$=2--1 at 121.697~$\mu$m \citep{Neu97}
and the o--H$_2^{18}$O  2$_{12}$--1$_{01}$ line 
at 181.053~$\mu$m has a small contribution from H$_3$O$^+$ Q(1,1) \citep{Goi01}.

Observations of the 183.31~GHz H$_2$O line (E$_u$$\simeq$205~K) 
toward Sgr~B2 are presented in Fig.~\ref{mapa_183}. 
The emission appears  at the LSR velocities of Sgr~B2
with no contribution from the line of sight clouds. The emission appears at 
least in the 50-75~km~s$^{-1}$ range. Note the different line shapes and 
intensities of the 183.31~GHz emission for positions in front and around 
the main condensations (here the line appears much wider).

Finally, Fig.~\ref{co_7_6_cso} shows the CO $J$=7--6 line (E$_u$$\simeq$150~K) 
observed toward Sgr~B2(M). This CO line shows a strong 
self--absorption at $sim$70~km~s$^{-1}$ (the velocity of the 183.31~GHz H$_2$O 
line peak at his position) so that the profile peaks   
at $\sim$ $+$55 and $+$85~km~s$^{-1}$. Similar patterns are shown by 
lower-J CO lines.

\section{Analysis and discussion}
\label{sct:analysis}

\subsection{Carbon Monoxide}
\label{ssct:co}
\begin{figure}[t]
\centering
\begin{center}
\includegraphics[angle=0, width=8cm]{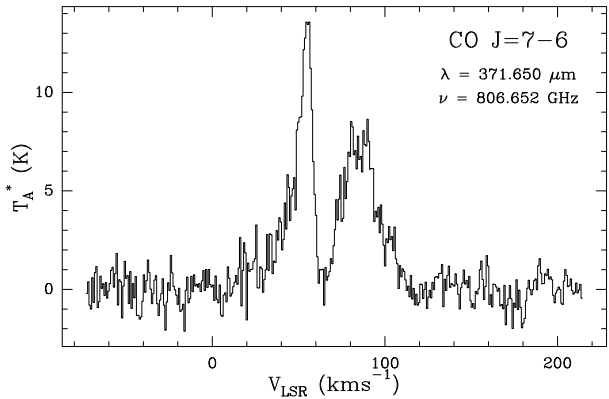}
\caption{CSO observation of the CO $J$=7--6 line at 806~GHz
toward Sgr~B2(M).}
\label{co_7_6_cso}
\end{center}
\end{figure}

The warm gas present in the outer layers of Sgr~B2(M) might be expected
to radiate in high--$J$ CO lines. The CO $J$=14--13 transition at 185.999~$\mu$m 
is the one with the lowest energy level (E$_u$$\simeq$581~K) within the range of 
ISO/LWS detectors.
However, we have not detected any emission/absorption from CO 
(3$\sigma$ limits are $<$2$\times$10$^{-18}$~W~cm$^{-2}$) 
in the ISO/LWS spectra 
toward the Sgr~B2 region \citep{G04}, with both grating and FP. 
CO spectroscopical data and line flux upper limits are tabulated in 
Table~\ref{tab-co_lines}. 
Nevertheless, recent  studies of the large scale CO $J$=7--6 emission toward the 
GC with the \textit{Antarctic Submillimeter Telescope and Remote Observatory} (AST/RO)
have shown that the emission is concentrated toward Sgr~B and Sgr~A
complexes \citep{Kim02}. Fig.~\ref{co_7_6_cso} shows higher spectral/angular resolution 
observations of this line taken with the CSO telescope toward the Sgr~B2(M) position.
Hence, it seems that at a given $J_{up}$ level, the rotational CO line emission 
disappears from the Sgr~B2 spectrum. 

\begin{figure*}[t]
\centering
\begin{center}
\includegraphics[width=8cm,angle=-90]{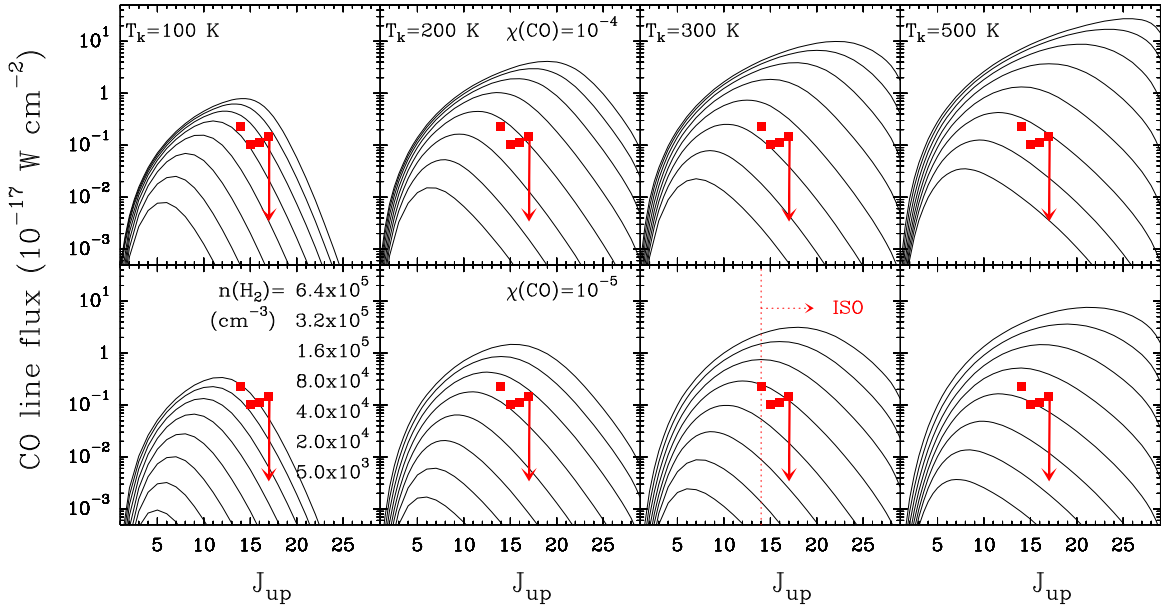}
\caption{Nonlocal models of CO rotational
lines toward Sgr~B2. 
Velocity--integrated line fluxes (in 10$^{-17}$ W cm$^{-2}$) are shown as a function
of $J_{up}$, the CO upper level rotational number.
From $J$=14--13,  CO lines are  accessible to the
ISO/LWS (vertical dashed line). No CO line has been detected.
3$\sigma$ upper limits for CO line fluxes are shown (filled squares).
Two different abundances have been considered, $\chi$(CO)=10$^{-4}$ 
(upper panels) and $\chi$(CO)=10$^{-5}$ (lower panels). 
Models with different kinetic
temperatures (100 to 500~K) are shown in each panel for
different H$_2$ volume densities: 5.0$\times$10$^3$, 
1.0$\times$10$^4$, 2.0$\times$10$^4$, 4.0$\times$10$^4$, 8.0$\times$10$^4$, 1.6$\times$10$^5$, 
3.2$\times$10$^5$, 3.2$\times$10$^5$ and 6.4$\times$10$^5$~cm$^{-3}$.}
\label{co_models}
\end{center}
\end{figure*}


We have performed nonlocal radiative transfer calculations to try to reproduce 
the lack of high--$J$ CO lines in the
far--IR spectrum of Sgr~B2(M), and to help constraining the physical parameters
needed to model the water vapor absorption. In Fig.~\ref{co_models} we 
show the predictions of a nonlocal model for several
high--$J$ transitions of CO. The nonlocal model used here
is an adaptation of the radiative transfer code developed by 
Gonz\'alez--Alfonso \& Cernicharo \citep{Gon93}
with the inclusion of dust in the transfer \citep{Cer00}.
We have implemented it for 
high--$J$ levels of CO (see Sec. \ref{sssct:nonlocal} for further details on the specific model
developed for Sgr~B2(M)).
The collisional rates have been taken from Flower (2001). The dust continuum emission has 
been modeled following Goicoechea and Cernicharo (2002) and G04 (see section 4.2.1).
We have run an array of models for CO abundances\altaffilmark{5} of 10$^{-4}$ and
10$^{-5}$ with T$_k$ varied from 100 to 500~K
and $n(H_2)$ from 5.0$\times$10$^3$ to 6.4$\times$10$^5$~cm$^{-3}$. From these 
results it is clear that in order to suppress the far--IR CO emission, 
and to match the CO $J$=7--6 line emission, low H$_2$ density is required.  
In particular, if the CO $J$=7--6 line arises from a layer of gas
at T$_k$$\sim$100~K the model with $n(H_2)$=2$\times$10$^4$~cm$^{-3}$ correctly
reproduces the absence of far--IR lines. However, if it arises from
the same outer layer of warm OH (T$_k$$\sim$300~K) detected 
in the far--IR \citep{Goi02}, the
limit for the density will be $<$10$^4$~cm$^{-3}$.
In any case, the lack of high--$J$ CO lines is an  evidence of 
the low density gas located in front of the far--IR continuum source.
In the following section we will constrain the physical parameters
of this layer by studying the available water vapor lines.

\footnotetext[5]{An upper limit to the  $^{13}$CO fractional abundance in
Sgr~B2 of 10$^{-6}$ was found by Lis \& Goldsmith \citep{Lis89}. According to
$^{12}$C$^{18}$O/$^{13}$C$^{18}$O$\simeq$25 in Sgr~B2 
\citep{Lan90}, the $^{12}$CO abundance would be 
$\lesssim$2.5$\times$10$^{-5}$.}

\subsection{Water Vapor}
\label{ssct:h2o}
The main observational result of this work is that the far--IR water lines
toward Sgr~B2(M) appear in absorption, while the 183.31~GHz line is seen in
emission around and in the main condensations. The fact that the continuum emission
in the far--IR is optically thick \citep{G04} indicates that the
H$_2$O absorption lines arise from regions where the excitation temperatures
(T$_{ex}$) are smaller than the dust temperatures 
inferred from the continuum emission.
However, the problem of the H$_2$O  line excitation toward Sgr~B2
is not straightforward.
Apart from self--absorption and the possible excitation by collisions 
with molecules (e.g. H$_2$), atoms (e.g. He) and $e^-$ (if the ionization
fraction is significant), the level population can be primarily
determined by the thermal emission of dust, the role of which 
is essential in the excitation of molecules such as H$_2$O
or OH which have many rotational lines in the far-- and mid--IR.
This represents a major difference with respect to the excitation treatment of  
molecules observed in the radio domain where, generally, one can neglect
the excitation by dust photons.
If the excitation is dominated by IR photons, the two transitions arising
from the ground levels of ortho-water, 
2$_{12}$--1$_{01}$ and 1$_{10}$--1$_{01}$,  
will determine how the higher energy levels 
will be populated.
Even if collisions are important, the presence
of an optically thick far--IR continuum 
will strongly affect the T$_{ex}$ of water 
and, thus,
it must be carefully taken into account in the models.
In the case of Sgr~B2, this means that
the external dust layers of the cloud will absorb 
the possible water line emission from the inner regions.
Knowledge of the geometry of the region to be modeled and the
relative filling factors of the dust and gas in the beam of the LWS instrument, 
are also important for the models. For this reason, high angular resolution 
ground-based observations of the 183.31 GHz water line are particularly important 
to model the water vapor radiative transfer in Sgr~B2.


\subsubsection{Large Velocity Gradient modeling}
\label{sssct:lvg}
\begin{figure}[t]

\includegraphics[angle=0, width=7cm]{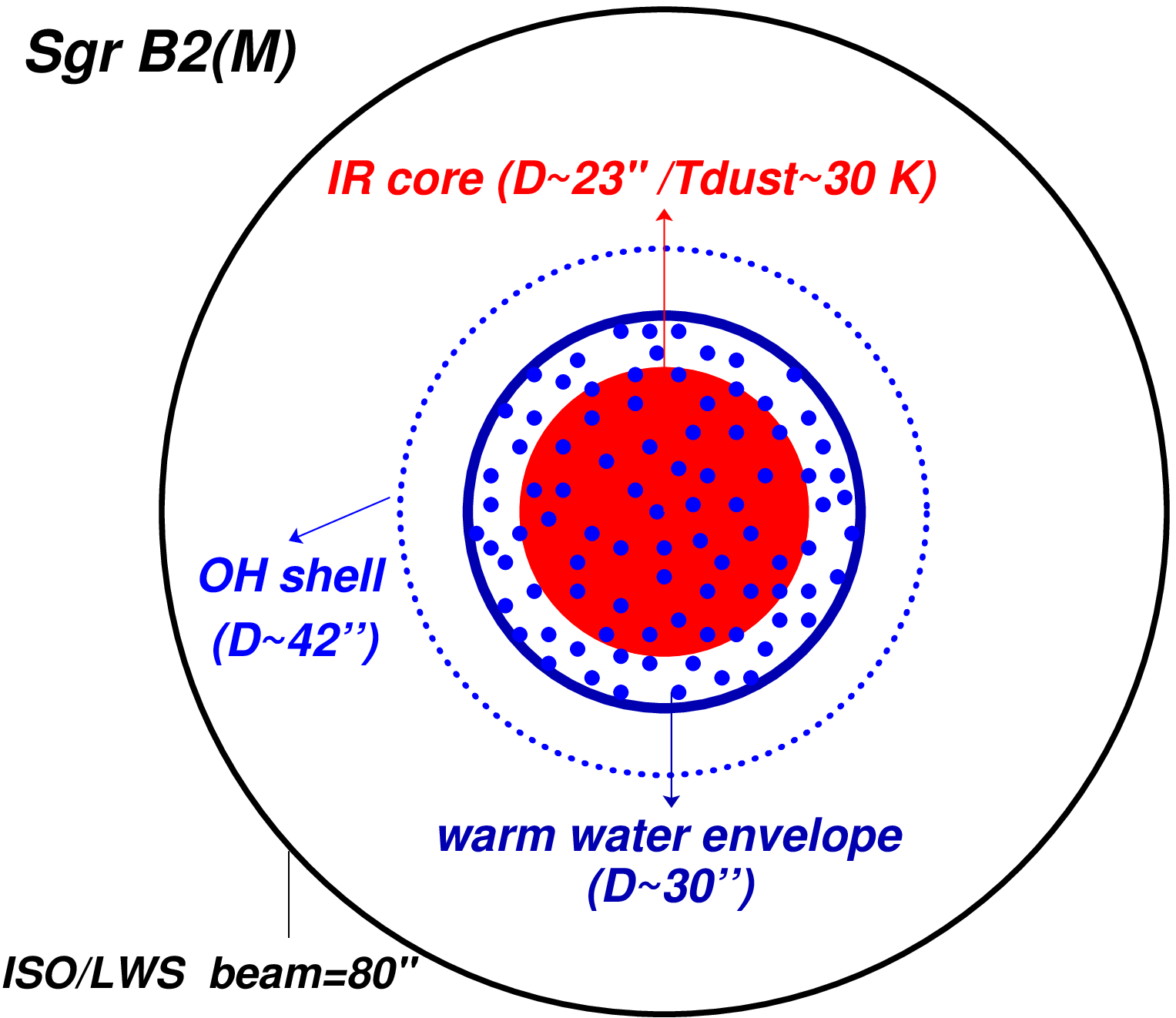}
\centering
\caption{Sketch of the water model. The dimensions of the OH
envelope are also shown (Goicoechea \& Cernicharo 2002).}
\label{fig_h2o_geometry}
\end{figure}

In this section we analyze the H$_2$O observations with a 
\textit{multi-molecule Large Velocity Gradient} (LVG) model.  
For Sgr~B2(M) we have adopted a spherical geometry with 
two components (see Fig.~\ref{fig_h2o_geometry}): a uniform continuum core
with a diameter of $\sim$23$''$ ($\sim$1~pc for a distance 
of 8.5~kpc) and a shell of variable thickness and distance to the core.
The presence of an external shell of molecular gas (not resolved by the 
ISO/LWS beam) surrounding  a central condensation is indicated by the 
analysis of  the far--IR OH \citep{Goi02} and NH$_3$ lines 
\citep{Cec02} and it is also supported by H$_2$O 183.31~GHz 
line observations (Fig.~\ref{mapa_183}).
In particular Goicoechea \& Cernicharo (2002) found an angular size of
$\sim$42$''$ for the OH envelope. The bulk of the H$_2$O absorption
lines must arise from the OH layers, or inside them. Following the OH model
geometry and Figs.~\ref{mapa_183} and \ref{co_7_6_cso}, a total size of
$\sim$30$''$ ($\sim$1.2~pc) for the core$+$shell cloud has been adopted.
The core is considered as a gray--body with an opacity at 80~$\mu$m of 2.5,
with a dust opacity law given by $\tau_{\lambda}=\tau_{80}[80/\lambda(\mu m)]$,
and a dust temperature of 30~K. 
These values are consistent with the color temperatures and
dust emissivities derived from
the analysis of the ISO/LWS continuum observations at the same
wavelengths of the detected far--IR H$_2$O lines \citep{G04}.
Although dust temperatures can be slightly larger (T$_d$$\sim$60--80~K) 
inside the cloud
(from the analysis of the millimeter continuum emission) or slightly
lower (T$_d$$\sim$20~K) in the outer and colder diffuse layers 
(from the analysis of extended IRAS continuum emission; Gordon et al. 1993),
we judged T$_d$$\simeq$30~K as the most representative value for the dust
grains coexisting with the molecular species detected in the far--IR. The 
bulk of the continuum emission detected by ISO can be hardly fitted with 
dust temperatures below 30 K. The emission of higher temperature dust arising 
from the innermost regions of Sgr B2 is hidden in the FIR by the huge amount 
of foregroung gas and colder dust.

For these models we have considered
an ortho--H$_2$O column density of 1.8$\times$10$^{16}$~cm$^{-2}$
and a para--H$_2$O column density of 0.6$\times$10$^{16}$~cm$^{-2}$.
These are lower limits suggested by the nonlocal radiative transfer models (see below).
The LVG model computes the statistical equilibrium population of the
rotational levels for ortho--H$_2$O and para--H$_2$O independently.
The collisional rates were scaled from those of H$_2$O--He collisions
\citep{Gre93}.


\begin{figure*}[]
\centering
\includegraphics[ width=13cm]{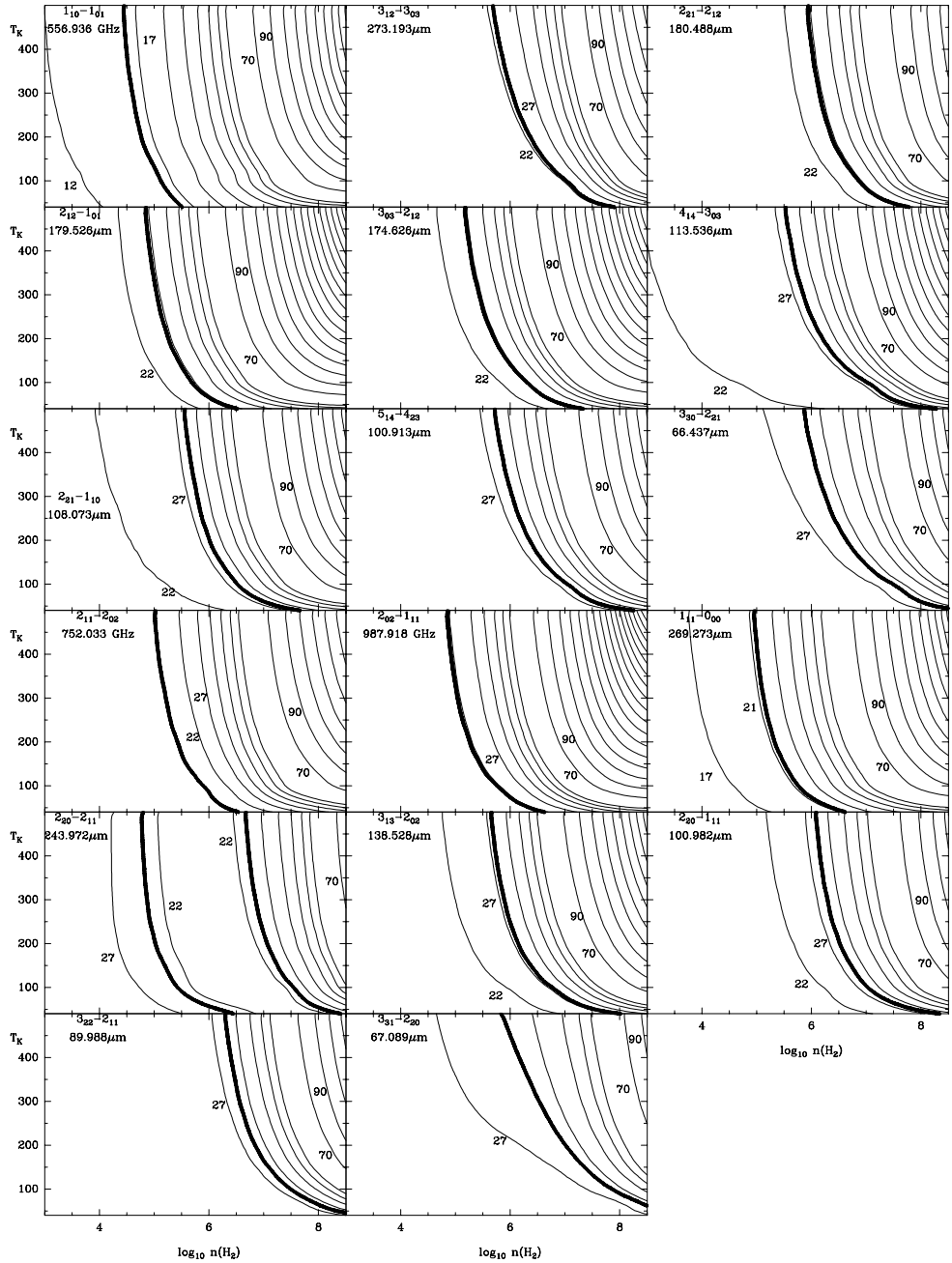}
\caption{Large Velocity Gradient excitation models of different kinetic 
temperature and density for several water vapor rotational lines.
The column density of ortho--H$_2$O is 1.8$\times$10$^{16}$~cm$^{-2}$ and that
of para--H$_2$O is 0.6$\times$10$^{16}$~cm$^{-2}$.
In each panel, the H$_2$O transition and wavelength$/$frequency is labeled. 
The thick contour corresponds to the equivalent temperature of the continuum
source (T$_c$). Hence, to the left of this contour, lines will be in absorption
(T$_{ex}$$<$T$_c$).}
\label{lvg_mod}
\end{figure*}

\begin{figure*}[]
\centering
\includegraphics[width=17cm]{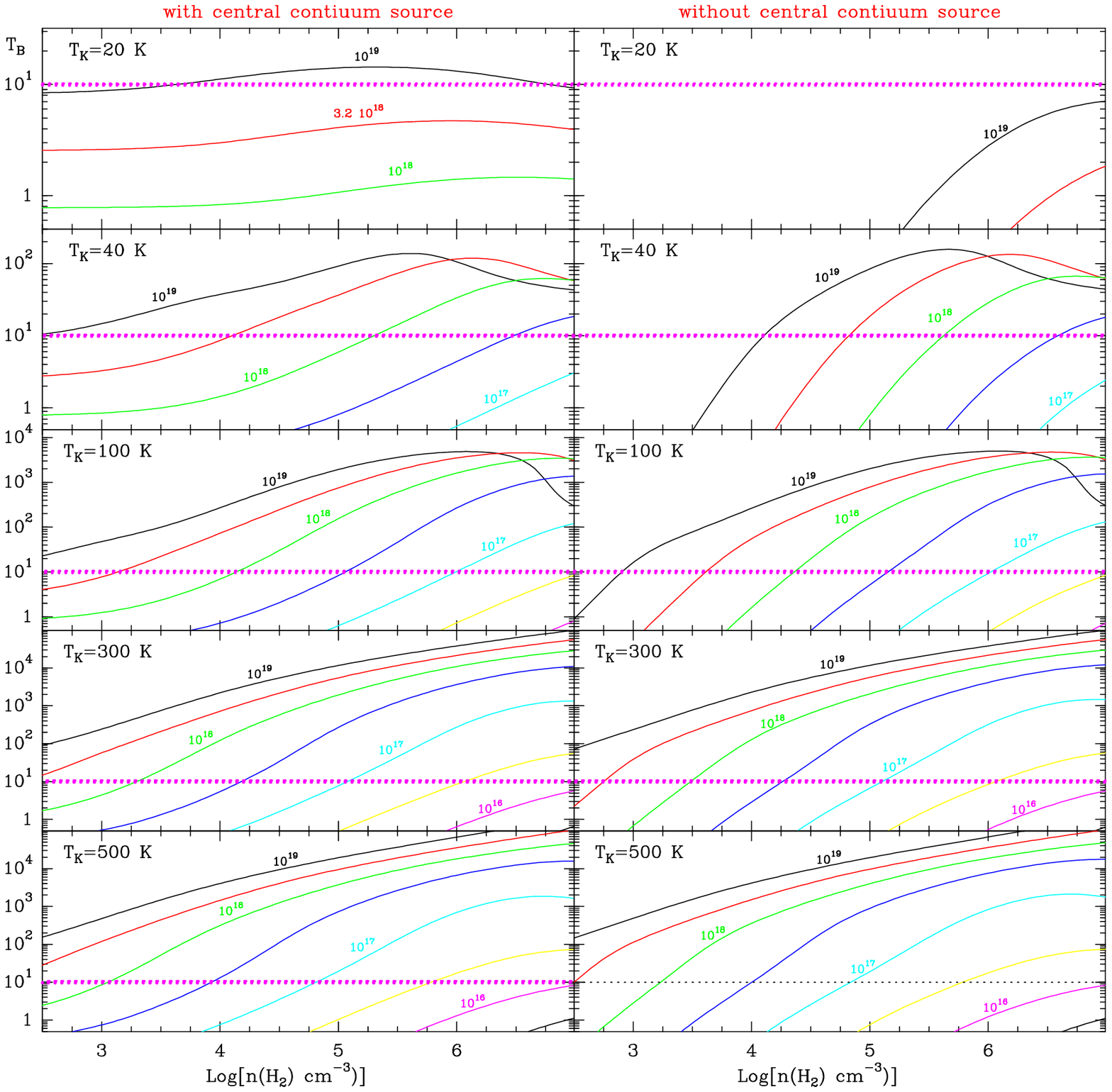}
\caption{Large Velocity Gradient excitation models for the 3$_{13}$--2$_{20}$
maser transition of para--H$_{2}^{16}$O at 183.31~GHz. Models with and without
a central continuum source (described as a gray--body with an 
opacity of 2.5 at 80~$\mu$m, with a dust opacity law given 
by $\tau_{\lambda}=\tau_{80}[80/\lambda(\mu m)]$, and a dust temperature 
of 30~K) have been considered. For low temperature and low densities, infrared
pumping seems to be efficient to increase the
emerging flux of the 183.3 GHz line. In addition
the maser amplifies the continuum. Hence, the resulting
flux depends on the assumed opacity at 80 $\mu$m and on
the opacity law. For higher kinetic temperatures and
densities collisional pumping of the maser
dominates and the difference between the two cases,
with and without central continuum source,
is less important. For a given kinetic
temperature (20 to 500~K), the expected T$_B$ as a function of the H$_2$
density is shown in each panel for different column densities of para--H$_{2}^{16}$O.
The dashed line indicates the observed brightness temperature in the extended emission. 
Toward the main condensations, brightness temperatures of $\sim$100 K are reached.}
\label{lvg_mod_183}
\end{figure*}

Several LVG computations for some selected ortho-- and para--H$_2$O transitions
in the THz domain are shown in Fig.~\ref{lvg_mod}. 
The excitation temperature T$_{ex}$ of each transition is shown in each panel
as a function of T$_k$ and $n(H_2)$.
The thick contour corresponds to the equivalent temperature of the continuum
core. Therefore, to the left of this contour, water lines appear 
in absorption. From the LVG models it is clear that
high density and temperature are required to observe the far--IR water lines
in emission. For the range of densities implied by the CO
observations ($<$10$^4$~cm$^{-3}$), the  H$_2$O lines observed by the ISO/LWS
are correctly predicted in absorption. The temperature of the absorbing 
layer is, however, more difficult to estimate because  the possible 
solutions for a given (low) density model are not very sensitive to  temperature
variations, as indicated by the smooth change of T$_{ex}$ for constant density 
as the kinetic temperature goes from 500 to 100 K.
LVG models predict that, in addition to dust photons, collisions play a role in 
the excitation of the lowest H$_2$O rotational levels.  
The higher energy levels are pumped from the lowest levels by 
absorption of far--IR photons. In the case of the ortho--H$_2$O 
1$_{10}$--1$_{01}$ line observed by \textit{SWAS} (Neufeld et al. 2000), 
only moderate densities ($>$5$\times$10$^4$~cm$^{-3}$)
are needed to observe the line in emission. This is the case of the
extended  emission in the 1$_{10}$--1$_{01}$ line observed in the \textit{180~pc
Molecular Ring} around the GC 
between $v_{LSR}$$\sim$$+$80 and 120~km~s$^{-1}$
(Neufeld et al. 2003). For the density conditions derived for the
Sgr~B2 velocity range and for the line of sight clouds, the 557~GHz line is
correctly predicted in absorption.
A similar behavior is expected for the para--H$_2$O 1$_{11}$--1$_{00}$ line
at $\sim$269.3~$\mu$m ($\sim$1113~GHz)  that will be observed 
by  future heterodyne instruments such as HIFI/Herschel.

Among all the moderate excitation  para--H$_2$O lines (E$_l$$<$450~K), only the  
3$_{13}$--2$_{20}$ appears in the mm domain. Contrary to other H$_2$O maser 
lines accessible from ground--based
telescopes, relatively low T$_k$ and  density are required for the
3$_{13}$--2$_{20}$ line inversion. These conditions allowed the first detection
of extended water emission in Orion (Cernicharo et al. 1994), and  
a strong dependence of the emission with T$_k$ was revealed.
Hence, the 183.31~GHz 
line could be an excellent tracer of the warm gas in molecular clouds. 



We have used the same LVG model to analyze the 183.31~GHz line 
inversion mechanism
that produces  extended emission in Sgr~B2 (Fig.~\ref{mapa_183}). 
Fig.~\ref{lvg_mod_183} shows different
excitation models  for the 3$_{13}$--2$_{20}$ transition. 
Since a fraction of the 183.31~GHz line emission may arise from
the warm envelope in front of the far--IR continuum source,
models with and without
a central continuum source (the same described at the beginning of the section)
have been considered. For a given T$_k$ (from 20 to 500~K),
each panel shows computations of different para--H$_2$O column densities as 
a function of  T$_B$ (in K) and  H$_2$ volume density.
For low temperatures (T$_k$$<$40~K) and moderate densities
($<$10$^5$~cm$^{-3}$), 
the line will be observable for large $N$(p--H$_2$O) values only if a
continuum source is present. Due to the minor role played by collisions in the
pumping of the rotational levels at these temperatures, the line intensity 
is almost independent of 
the density. Nevertheless, as the temperature increases, collisions 
start to be significant, and the expected intensity of the 183.31~GHz
line becomes less sensitive to the models with or without the continuum
source.

The 183.31~GHz emission is produced by the population inversion of the 3$_{13}$
and 2$_{20}$ levels. This is due to the different rates
(far from the thermalization densities) at which both levels can be populated.
The 3$_{13}$ level is radiatively connected with the 2$_{20}$ and
2$_{02}$  lower energy levels. The A$_{ij}$ coefficient for the 3$_{13}$--2$_{02}$
at $\sim$138.5~$\mu$m is $\sim$10$^4$ times larger than the A$_{ij}$ of the 
3$_{13}$--2$_{20}$ transition.
The $\sim$138.5~$\mu$m 
line is seen in strong absorption in the ISO/LWS spectra (see Fig.~\ref{obs_iso_agua}).
The 2$_{20}$ level is radiatively connected with the 2$_{11}$ and  
1$_{11}$ lower energy levels. The 2$_{20}$--1$_{11}$ at $\sim$100.9~$\mu$m
produces the strongest absorption of para--H$_2$O in the far--IR 
(see Fig.~\ref{obs_iso_agua}).
On the other hand, the 3$_{13}$ level is radiatively connected with the 
3$_{22}$, 4$_{22}$ and 4$_{04}$ higher energy levels, while the 2$_{20}$
level is connected with the 3$_{31}$ (see the $\sim$67.1~$\mu$m line in
Fig.~\ref{obs_iso_agua}) and with the 3$_{13}$ higher energy levels.
The different radiative pathways and rates
(for $n$$<<$$n_{cr}$) at which the 3$_{13}$
and 2$_{20}$ levels can be populated  produce the inversion.
LVG models (see Fig.~\ref{lvg_mod_183}) show that this mechanism can be efficient
to produce 183.31~GHz line emission in regions of relatively low density.

The observed extended emission at 183.31~GHz has a brightness temperature of 
10 K in average (see Fig.~\ref{mapa_183}). Assuming that it arises from the 
low density regions in which the the far--IR continuum sources are embedded, 
with T$_k$$=$300-500 K and $n(H_2)$$\sim$10$^4$~cm$^{-3}$, 
we derive a value for $N$(p--H$_2$O) of $\gtrsim$5$\times$10$^{17}$~cm$^{-2}$. 
This  value is higher than the one derived from the far-IR water lines (see sect. 
\ref{sssct:nonlocal}) possibly because 
the 183.31 GHz line penetrates deeper into this dusty environment. Toward the main 
condensation, the bulk of the emission seems to arise from the cold (T$_k\sim$40 K) 
and dense gas. Under these conditions, a brightness temperature of the line of $\sim$100 K 
would translate into a  p--H$_2$O column density of $\sim$10$^{19}$~cm$^{-2}$. However, an 
important contribution could come from the embedded  high temperature and high density 
core condensations where the column density should also be much larger.

\subsubsection{Non-Local radiative transfer models}
\label{sssct:nonlocal}
\begin{figure*}[t]

\includegraphics[angle=0, width=12.7cm]{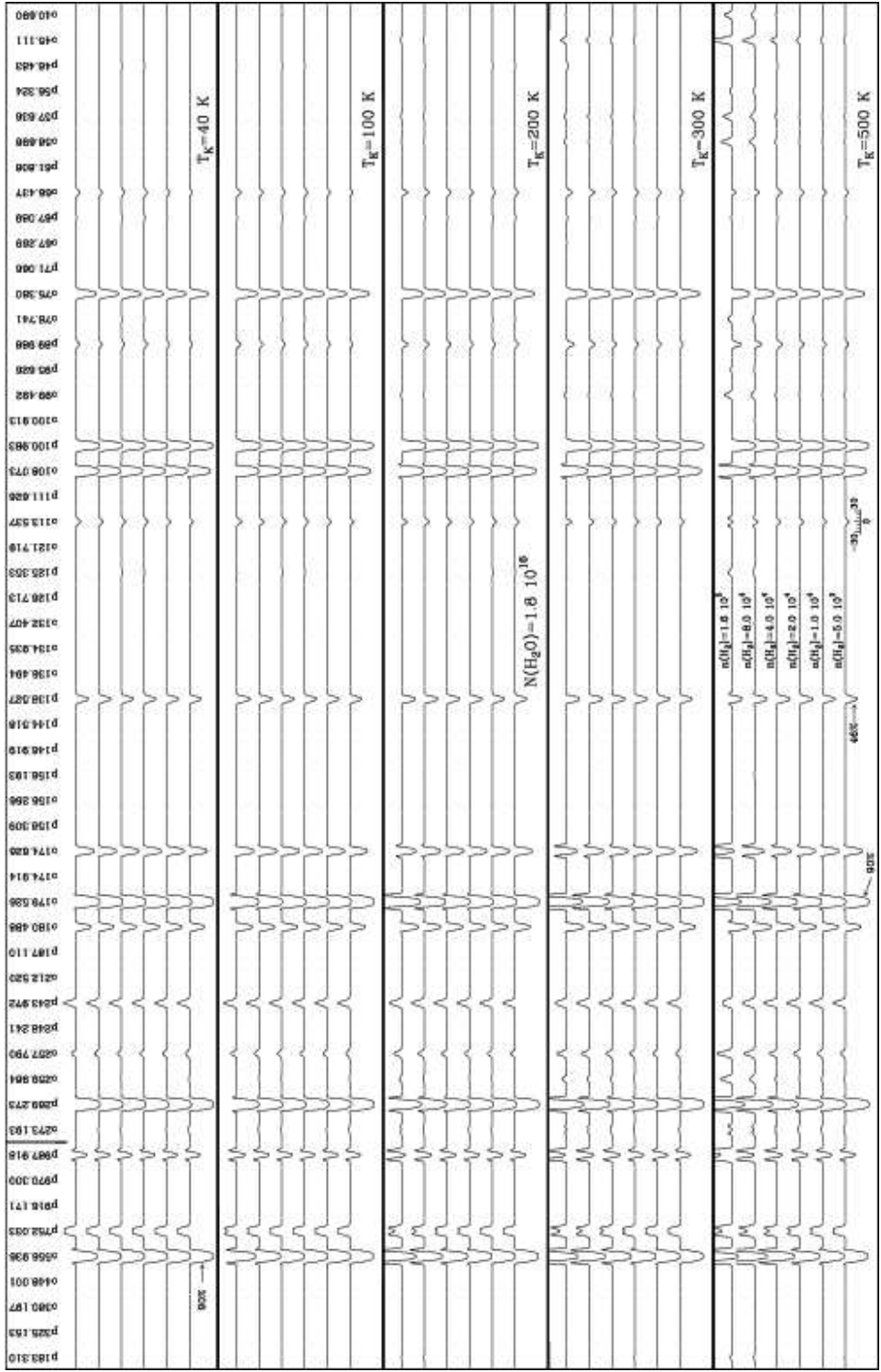}
\centering
\caption{Results from selected nonlocal models of H$_2$O rotational
lines in Sgr~B2(M). Results for different gas temperatures
(from 40 to 500~K) and H$_2$ densities (from 5.0$\times$10$^3$ to
1.6$\times$10$^5$~cm$^{-3}$) are shown. Note that for each model, the water 
abundance changes accordingly with n(H$_2$) in order to keep constant the water 
vapor column density (in this case, 1.8$\times$10$^{16}$~cm$^{-2}$).  
For the first nine lines, the frequency is given in GHz; for the rest, the 
wavelength in $\mu$m is indicated. The velocity scale is shown at the bottom of the 
113.5 $\mu$m line.}
\label{mod_1_h2o}
\end{figure*}

\begin{figure*}[h]
\centering
\includegraphics[angle=0, width=13.0cm]{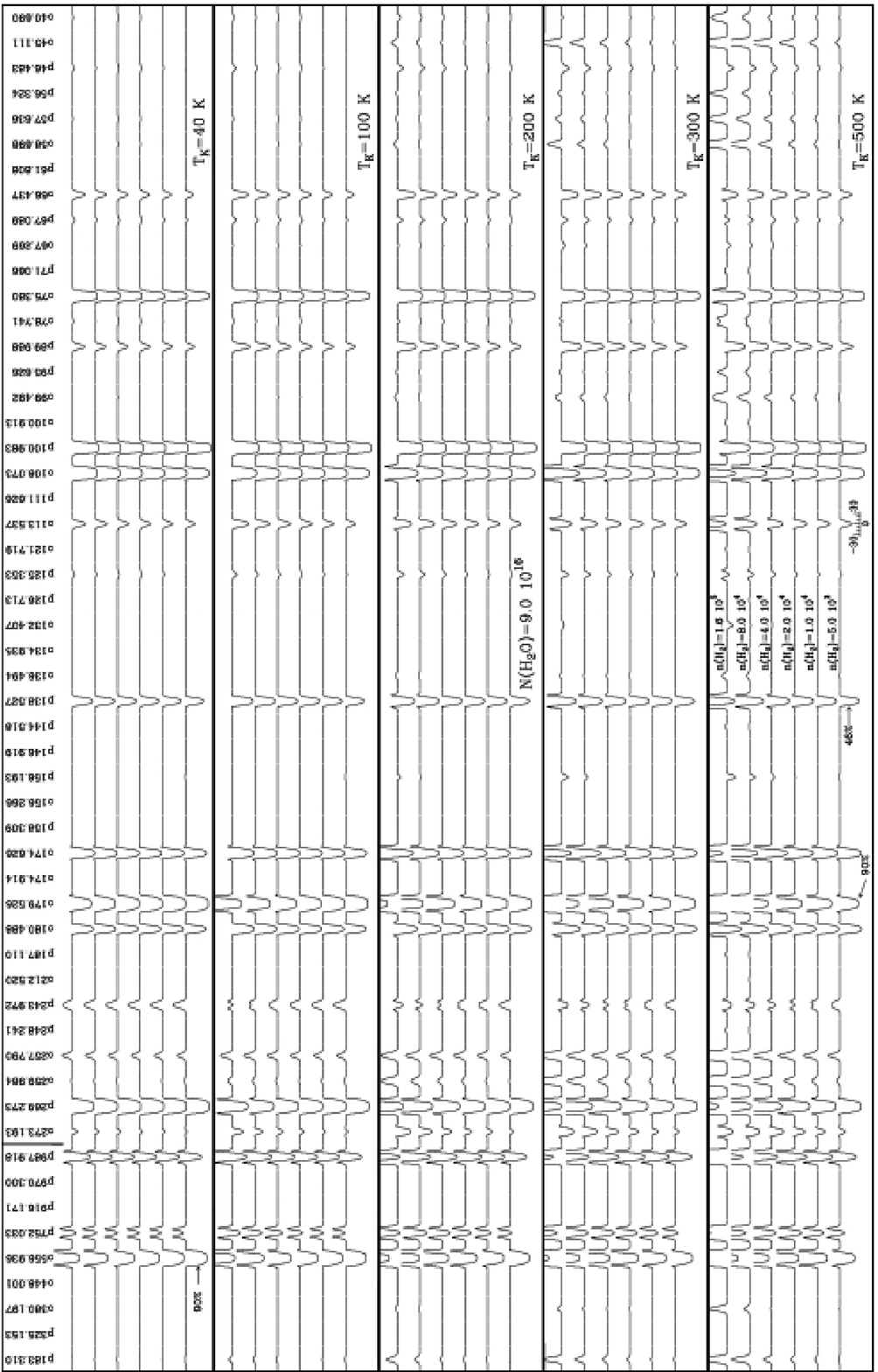}
\caption{As in Fig.~\ref{mod_1_h2o} but for
a water vapor column density of 9.0$\times$10$^{16}$~cm$^{-2}$.}
\label{mod_2_h2o}
\end{figure*}

To take into account the  radiative coupling between regions with
different physical and/or excitation conditions, the radiative transfer has to be treated
with nonlocal techniques, more sophisticated than the LVG approximation.
We have adapted the radiative transfer used in Sec.~5.1 for CO, to the 
ground vibrational states of
ortho-- and para--H$_2$O respectively. The model includes all the water rotational
levels with transitions between 40~$\mu$m and 183.31~GHz. The level population
is computed in statistical equilibrium considering  collisional excitation
and  radiative excitation by line and continuum photons.  This is computed consistently
assuming that the water molecules and the dust grains are coexistent.
The geometry, core$+$shell dimensions, are the same of that considered in the LVG 
models.
The shell was divided in 41 layers. The central dust condensation has been modeled 
with identical parameters than in section 4.2.1. We have considered 14 rotational levels (E$_{u}$ $<$ 608 K) 
of ortho-water for model with T$_k$ below 100 K , and up to 30 rotational levels (E$_{up}$ $<$ 1290 K) for the 
higher temperature models. We have adopted a turbulence velocity of 8 km~s$^{-1}$. 
The continuum radiation field has been treated 
as in Gonz\'alez-Alfonso \& Cernicharo \citep{Gon93} by considering a spectral range of 70 km~s$^{-1}$ 
centered on 
each water transition. Collisional rates have been taken from 
Green et al. (1993). 
Obviously all the water transitions in the continuum core are
thermalized to the dust temperature due to the large opacity in the far--IR.
The computed line profiles are a result of convolving the brightness temperature 
with the angular resolution
of the LWS detectors ($\sim$80$''$). The resulting spectral resolution of the
synthetic water lines is 1~km~s$^{-1}$, as no convolution with the
spectral resolution has been performed.
To test the sensitivity of the model to the physical parameters, models were 
computed with $N$(H$_2$O) and T$_k$ of 
1.8$\times$10$^{16}$, 9$\times$10$^{16}$ and
4.5$\times$10$^{17}$~cm~$^{-2}$, and
40, 100, 200, 300 and 500~K respectively, while densities were increased
from 5$\times$10$^{3}$ to  6.4$\times$10$^{5}$~cm~$^{-3}$ multiplying by 2 
in each step. The different nonlocal radiative transfer models 
for the first two H$_2$O column 
densities are shown in Figs.~\ref{mod_1_h2o} and \ref{mod_2_h2o}. Models 
for other column densities have been also ran. However, the observed absorption depth 
is not reproduced for column densities below 10$^{16}$ ~cm~$^{-2}$. In the 
$N$(H$_2$O)=1.8$\times$10$^{17}$~cm~$^{-2}$ case too many lines would be in 
emission, contrary to the observations.

The main problem to interpret the H$_2$O absorption toward Sgr~B2 arises from the
large opacities of the far--IR water lines, $\sim$10$^3$, and  even
$\sim$10$^4$ for the ortho--H$_2$O 2$_{12}$--1$_{01}$ line at $\sim$179.5~$\mu$m.
Under these conditions, many weak water lines have to be detected
to constrain the physical conditions and the column density. 
Therefore, we strength that it is  difficult to obtain physical parameters from the
observation of few far--IR water rotational lines. In addition, as the radiative excitation
dominates the population of the far--IR levels, lines with a weak dependence
on the dust excitation should be investigated. 
Taking into account the 14 water lines detected in the far--IR, 
the nonlocal results imply that  models are not very sensitive
to the temperature and that the only indication about the column density
has to be searched in weak far--IR H$_2$O lines or in the 
183.31~GHz line (see Sec.~6.1.2). Even so, the far--IR absorption  arises 
in the low density external layers of gas, while it is very likely that
the 183.31~GHz line could have an important contribution from inner and 
denser regions.
Another complication arises from the fact that for models of low $N$(H$_2$O)
values and T$_k$$\lesssim$200~K, it is also difficult to distinguish 
between different H$_2$ densities (see Fig.~\ref{mod_1_h2o}). Limits to
T$_k$ and $n$(H$_2$) have to be searched in weak  lines below 70~$\mu$m.

The models for high water column densities (Fig.~\ref{mod_2_h2o})  
predict 
absorption lines in the LWS range  at $\sim$56.3, $\sim$57.6, $\sim$58.7, $\sim$78.7,
$\sim$99.5, $\sim$125.4 and $\sim$136.5~$\mu$m.
At the spectral resolution and sensitivity of the LWS/FP, none of these lines
has been detected. This implies that $N$(H$_2$O)$\le$4.5$\cdot$10$^{17}$~cm$^{-2}$ toward
the warm envelope of Sgr~B2. The $\sim$136.5~$\mu$m line (also predicted
by the models with large column densities) is contaminated by the absorption produced
by the C$_3$ $Q$(8) rovibrational line, which is also predicted by the
models of tri--atomic carbon \cite{Cer00}. 
Some of these lines are predicted (even in emission) by the  models with large
$N$(H$_2$O).
Models with $N$(H$_2$O)$=$1.8$\times$10$^{16}$~cm$^{-2}$
are consistent with  ISO detections and upper limits to other ISO lines.
Only the $\sim$67.1~$\mu$m line is weaker than the observations.
Therefore, models shown in Fig.~\ref{mod_1_h2o} give a lower limit to the 
water vapor column
density in the outer and warm (300-500 K) envelope.
Taking into account the difficulties implied by the H$_2$O modeling
in the far--IR, we found that $(9\pm3)\times10^{16}$~cm$^{-2}$  
is the best H$_2$O column density to fit the ISO observations (see Figure 
\ref{mod_2_h2o}).
The CO analysis (Sec.~5), and the studies in the far--IR of 
OH \citep{Goi02} and the ammonia lines \citep{Cec02} and our CO data, 
also support that the water 
absorption lines arise in the warm and low density ($\lesssim$10$^4$~cm$^{-3}$) 
layer in front of Sgr~B2(M). The H$_2$O column density derived from ISO observations 
is below the lower 
limit of $N$(H$_2$O) estimated for the 183.31~GHz line for this component 
(see Sec. \ref{sssct:lvg}) as it is likely that an important fraction 
of the 183.31~GHz emission arises from the inner and denser regions closer to 
Sgr~B2 main cores or even from them, as pointed out above. 
This component, with a mean H$_2$ density of $>$10$^{5-6}$~cm$^{-3}$, has been
traced by the NH$_3$ non--metastable emission lines, T$_k$$\simeq$100~K,
\citep{Hut93} and by the emission of HC$_3$N rotational lines,
T$_k$$\simeq$20--40~K, \citep{Lis91}.
The determination of the temperature  from mm emission lines 
is also complicated in these regions completely obscured to  ISO
observations. Therefore, the observed differences in the 3$_{13}$--2$_{02}$ 
line--intensity and --shape can be a combination of $N$(H$_2$O) and/or T$_k$
variations across the region.

The results presented in this section have been compared to those obtained from another radiative
transfer based on a different approach (Asensio Ramos \& Trujillo Bueno 2003; 
Asensio Ramos \& Trujillo Bueno 2006). This code is a generalization to spherical geometry of the very fast 
iterative Multilevel Gauss-Seidel (MUGA) and Multilevel Successive Overrelaxation (MUSOR) methods 
developed by Trujillo Bueno \& Fabiani Bendicho (1995) for the case of a two-level atom and generalized
by Fabiani Bendicho, Trujillo Bueno \& Auer (1997) to multilevel atoms in Cartesian geometries. The code also 
allows the application of the 
standard Multilevel Accelerated $\Lambda$-iteration (MALI) (Olson, Auer \& Buchler 1986). The angular information for
the calculation of the mean intensity is obtained by solving the radiative transfer equation 
along its characteristic curves (straight trajectories with constant impact parameter) with the aid of the 
short-characteristics formal solver with parabolic precision (Kunasz \& Auer 1988) . The statistical equilibrium 
equations are linearized with the aid of the preconditioning scheme developed by Rybicki \& Hummer (1991, 1992) with 
the introduction of an approximate $\Lambda$ operator that can be efficiently obtained in the framework of the 
short-characteristics technique. 
The convergence rate of the MUGA and MUSOR schemes is equivalent to that obtained with the introduction of 
a nonlinear $\Lambda$ operator, with the advantage of not being necessary neither to calculate nor to invert such nonlinear
operator. Interestingly, the time per iteration is similar to that obtained for the standard $\Lambda$-iteration
or the MALI method. The computing time for the MUGA scheme is reduced by a factor 4 with respect to MALI, while the
MUSOR scheme leads to an order of magnitude of improvement in the total computing time with respect to MALI.

The calculations have been performed with the same geometry, the same molecular data (collisional and 
radiative transitions), and the same physical conditions of the previously described nonlocal code. The emerging line profiles from both codes are 
very similar, with differences below 2-3 \%. Both codes predict lines in absorption/emission for the same physical conditions, with identical 
spectral shapes in the cases where re-emission 
is found in the line wings. This test of consistency allows us to be very confident in the results presented in 
Figs.~\ref{mod_1_h2o} and \ref{mod_2_h2o} since they were obtained with numerical methods based on completely different 
approaches.

\section{The warm envelope: Shocks, PDRs or XDRs ?}
\label{sct:discussion}

The H$_2$O (OH) column densities derived from ISO observations in the warm
envelope are within an order of magnitude of the shock model
predictions for the same low density region \citep{Flo95}.
Nevertheless, $C$--shocks have been invoked to
explain the heating of large amounts of warm gas in the GC
and also in the warm envelope \citep{Wil82,Mar97}. 
In addition, the shock models 
of \citep{Flo95} correctly reproduce the observed column densities
of $N$--bearing species such as NH$_2$ and NH$_3$ that
could be ultimately related with the dust grain chemistry \citep{G04}.

A qualitative explanation came from the observation of large OH column densities
(H$_2$O/OH$\sim$2-4) in the warm envelope.
According to Goicoechea \& Cernicharo (2002), if a far--UV radiation 
field illuminates the outer regions of the cloud, water molecules could be 
photodissociated producing an enhancement of the OH abundance expected only 
from its formation through H$_3$O$^+$ dissociative recombination.
The presence of such an extended far--UV radiation field with
$G_0$$\sim$10$^{3}$-10$^{4}$ is  inferred
from the [\OIII], [\NIII], [\NII], [\CII], and [\OI] extended line emission
\citep{G04}. Therefore, an important contribution from H$_2$O
photodissociation may explain the observed H$_2$O/OH abundance
ratio in the external layers of the envelope.\\

\begin{figure}[t]
\centering
\includegraphics[angle=-90, width=7.5cm]{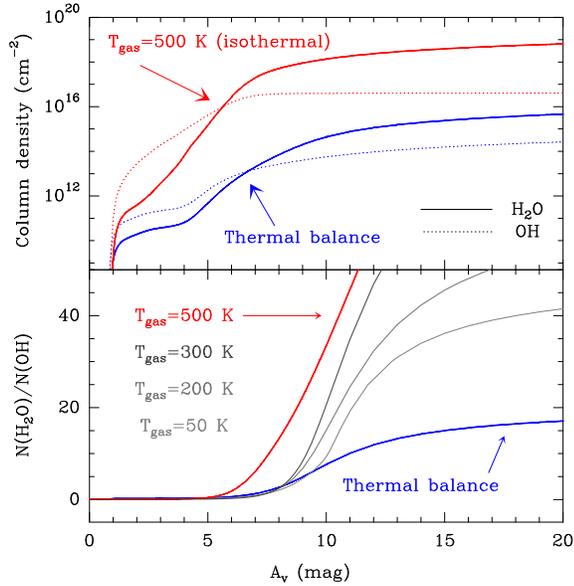}
\caption{Selected PDR models in which
$G_0$=5$\times$10$^3$ and $n_H$=5$\times$10$^3$~cm$^{-3}$.
\textit{Upper panel:} H$_2$O and OH column densities as a function of
the visual extinction for an isothermal model with T$_{gas}$=500~K and for
a model were the full thermal balance is solved at each depth.
\textit{Lower panel:} H$_2$O/OH column density ratio as a function of
the visual extinction for different isothermal models with
T$_{gas}$=500, 300, 200 and 50~K and for full thermal balance.}
\label{mods_pdr}
\end{figure}

To investigate in more detail the O--chemistry in the envelope, we have run several
photochemistry calculations that include depth--dependent photodissociation,
H$_3$O$^+$ dissociative recombination and neutral--neutral reactions.
We have used the latest version of the public and available PDR 
model\altaffilmark{6} of Le Bourlot et al. \citep{LBu93}.
\footnotetext[6]{http://aristote.obspm.fr/MIS/} 
The model does not include oxygen grain surface
chemistry, but it is consistent with the low densities
and high temperatures found in Sgr~B2 envelope.
Following \citep{G04}, we assume a $G_0$=5$\times$10$^{3}$ radiation field and a 
$n_H$=$n$(H)+2$n$(H$_2$)=5$\times$10$^{3}$ cm$^{-3}$ density.
Assuming that the outer gas layers of Sgr~B2 envelope are directly illuminated
by such a radiation field, the model solves the UV transfer and the chemistry 
up to A$_V$=20~mag. 
As a reference model we solve the thermal balance explicitly 
for an initial gas temperature of 500~K. The resulting H$_2$O and OH column
densities and the H$_2$O/OH ratio are shown in Fig.~\ref{mods_pdr}
as a function of the  visual extinction through the cloud.
This $pure$ PDR can only maintain the temperature above $\sim$100~K in the
first 2 magnitudes of the cloud, where the UV radiation efficiently 
photodissociates H$_2$O to form OH. This results in a low H$_2$O/OH ratio.
Inside the cloud, the H$_2$O and OH production is rapidly dominated
by  H$_3$O$^+$ dissociative recombination and the H$_2$O/OH ratio tends to
a constant value that basically depends on the assumed branching
ratio $f_{H_2O}$ (we have taken $f_{H_2O}$=0.25).

At least two different scenarios can produce large column densities
of gas at high T$_k$.
The first one is the presence of several clumpy PDRs within ISO's beam. 
In this scenario, far--IR H$_2$O and OH observations will only
trace the PDR--clump surfaces. 
The second one is the presence of shocks. It is very likely 
that widespread low velocity shocks can
locally heat the gas within the envelope and preserve a fraction of the molecular 
gas with temperatures up to $\sim$500~K \citep{Flo95}, but the same is possible in the 
clumpy PDR scenario. 
With these temperatures neutral-neutral reactions 
play an important role in the chemistry. 
To simulate this situation, we have run several iso--thermal models with temperatures 
ranging from 500 to 50~K. Some selected results are also shown in Fig.~\ref{mods_pdr}. 
From  these models it is clear that both $N$(H$_2$O) and $N$(OH)
are clearly enhanced by neutral-neutral reactions (Fig.~\ref{mods_pdr}, upper panel).
If the temperature is high, these models reproduce the observational 
values much better.  
Hence, a significant fraction of warm gas
seems to be  needed to reproduce the large H$_2$O and OH column densities observed 
toward Sgr~B2. This conclusion agrees with the large temperatures derived from
OH observations in the far--IR \citep{Goi02}.

The gas temperature determines the contribution of 
neutral-neutral reactions (1) and (2). As the temperature of the gas increases, 
reaction (1) contributes to a enhancement of the H$_2$O/OH ratio
(Fig.~\ref{mods_pdr}, lower panel). Note that a moderate increase of the  
radiation field (to $G_0$=10$^4$) shifts the H$_2$O/OH curve  to larger
A$_V$ inside the cloud until photodissociation becomes less important. 
In the shielded regions of the cloud, the
predicted column densities are similar to those with $G_0$=5$\times$10$^3$.

The H$_2$ column density (or A$_V$) responsible of the warm gas observed from
far--IR absorption lines (the location of the species in optical depth) 
has been traditionally  difficult to establish. 
A comparison between the observational H$_2$O/OH ratio and 
photochemistry models shows that the bulk of the 
H$_2$O/OH absorption must arise from the surface of Sgr~B2 
(a maximum A$_V$ of $\sim$5 to 10 mag), in agreement  with the large  opacities 
derived from the radiative transfer models. 
Even assuming an homogeneous surface cloud, most of the water vapor will
arise from A$_V$$<$10~mag if T$_k$$\sim$500~K while $\gtrsim$50$\%$ of water
can be at A$_V$$>$10~mag if T$_k$$\sim$300~K. Therefore, an accurate
description of the thermal structure of the cloud (with far--IR diagnostics
tracing the same gas) will be needed to establish more detailed conclusions.
Still, in a inhomogeneous medium, several PDR-like clump surfaces locally 
heated to
T$_k$$\sim$300--500~K by low velocity shocks could be entirely responsible 
for the
far--IR H$_2$O/OH absorption.

Taking into account the uncertainties implied in the determination of
column densities in the envelope, we take 
$\chi$(H$_2$O)$\simeq$(1-2)$\times$10$^{-5}$ as a lower limit. 
There is a determination of HD column density toward Sgr B2 of 
$\sim$10$^{18}$ cm$^{-2}$, which translates
into $N(H_2)\sim$10$^{24}$~cm$^{-2}$ (Polehampton et al. 2002).  
However, taking into account the huge absorption by dust 
at 112 $\mu$m ($\tau\sim$4), it is unlikely that in this HD column density the 
cold gas, where high column densities are expected, is accounted for.  We derive, 
see \citep{Goi02}, that the H$_2$ column density in the warm gas is 
$\sim$10$^{22}$ cm$^{-2}$. 
Besides, the 183.31~GHz observations imply that the water abundance is
at least an order of magnitude lower in the core regions..
However, we underline the importance of further detection of weak H$_2$O lines 
and/or lines with a lesser dependence from the dust emission to 
refine the models and better constrain the physical parameters of the region
(e.g. the temperature), that also determine much of the chemistry.
Specially important will be the input of the Herschel/HIFI observations
for water lines below $\sim$2~THz.\\

Another difficult  problem is to place the origin of the FUV radiation field
revealed by the fine structure line observations in the region \citep{G04}.
In principle, FUV photons could arise from the massive stars near the
Sgr~B2(M) core and/or from another stellar population within the envelope
itself (not resolved yet by observations). The permeating effect of the radiation
field will  be determined by the clumpyness and inhomogeneity of the medium
surrounding the stars, and by the energy of the stellar photons.
In addition, X--rays observations could 
complement this scenario of widespread low velocity shocks and UV radiation.
Energetic EUV-- and/or  X--ray photons  can penetrate deeper in the neutral cloud than
FUV--photons, and thus, could also play a role in the chemistry, as they can
also  induce many photo--ionization and photo--dissociation  processes.
The correlation found in the Sgr~B complex
between the 6.4~keV  Fe$^0$ line and the SiO emission could also 
indicate  that the X--ray sources  also drive the shocks in the region
\citep{Mar00}. 

Gas temperatures in XDRs can easily reach  T$_k$$\sim$300--500~K 
because of the more efficient gas heating by X--ray--induced photoelectrons
from the gas (and not from  dust grains as in PDRs). Under these
conditions, neutral--neutral reactions dominate 
the chemistry and especially OH reaches large abundances. However, 
XDR models predict low H$_2$O/OH$<$0.1 abundance ratios \citep{Mal96},
and thus, they can not be the dominant scenario explaining the far--IR
water and OH lines toward Sgr~B2 \citep{Goi02}.

Sgr~B2 shows diffuse emission in the K$\alpha$ line of Fe$^0$ at 6.4~keV
\citep{Mur01}. A  dozen of compact X--ray sources have been detected within Sgr~B2
cloud, and they may explain the whole X--ray emission found in the region.
Many of the detected X--ray sources are not associated with radio 
(\HII $\,$ regions created by massive stars) nor with IR sources. 
According to these observations, the intrinsic X--ray luminosity toward 
Sgr~B2(M) is $L_X$$\lesssim$10$^{35}$~erg~s$^{-1}$, which translates into 
a X--ray flux incident on molecular gas of $F_X$$\lesssim$0.1--0.001 erg~cm$^{-2}$~s$^{-1}$
(assuming $\sim$0.1--1~pc from the X--ray source). Hence, the X--ray field
within Sgr~B2(M) is in the low tail, or even weaker, than those studied by
Maloney et al. (1996).
As noted by G04, the low [\OI]63~$\mu$m/[\CII]158~$\mu$m and
([\OI]63+[\CII]158)/FIR intensity ratios observed in the region favor a dominant PDR
origin for these lines. Both shock \citep{Dra83}
and XDR \citep{Mal96} models predict larger ratios.

\section{Summary}
\label{sct:summary} 
We have carried out far--IR observations of several thermal absorption lines
of water vapor toward Sgr~B2(M) and have mapped the 183.31~GHz water line
around the main dust condensations of the complex. The main conclusions of
this work are the following:

\begin{enumerate}

\item The detected water absorption lines are very opaque and arise from 
the warm envelope around Sgr~B2(M). The observation of the CO $J$=7--6 line
and the lack of far--IR CO lines 
at ISO's sensitivities
(3$\sigma$ limits below $\sim$2$\times$10$^{-18}$~W~cm$^{-2}$)
imply that the density of such layer is 
$n$(H$_2$)$\sim$10$^4$~cm$^{-3}$.  The para--H$_2$O 3$_{13}$--2$_{02}$ line at 
183.31~GHz shows $\sim$40$''$$\times$40$''$ extended emission 
around Sgr~B2(M) and $\sim$40$''$$\times$20$''$
 around Sgr~B2(N). This is the first 
observation of that line in a GC source and represents further evidence that 
water vapor is extended in warm molecular clouds.

\item LVG and nonlocal radiative transfer 
calculations have been carried out to extract the water
vapor abundance and to constrain the physical parameters of the 
absorbing/emitting regions. Because of the radiative excitation by dust
photons, the far--IR water lines are not very sensitive to T$_k$. 
Taking into account the analysis of the related species OH \citep{Goi02},
the water absorption must arise from warm gas at similar 
temperatures, i.e., from 300 to 500~K.
For this warm envelope, we found $N$(H$_2$O)=$(9\pm3)\times10^{16}$~cm$^{-2}$.
An important fraction of the 183.31~GHz emission arises from the inner, denser and 
colder gas located closer to the main cores. We estimate a water abundance
of a few$\times$10$^{-7}$ in the denser regions.

\item Photochemistry models show that a component of warm gas,
$\sim$300-500~K, is needed to activate the 
neutral-neutral reactions and reproduce the large H$_2$O and OH column densities
observed in the envelope. We show that OH and H$_2$O far--IR observations
toward Sgr~B2 are surface tracers of the cloud (a maximum A$_V$ of 5 to 10 mag). 
We found $\chi$(H$_2$O)$\simeq$(1-2)$\times$10$^{-5}$ in these regions. 
Although irradiated by FUV, and possibly more energetic
photons, affecting the H$_2$O/OH ratio in the outermost layers, a 
clumpy structure for the PDR is needed. Alternatively, 
low velocity  shocks could maintain 
the gas heating through the envelope. 

\end{enumerate}

Due to the complexity of Sgr~B2 (and also of the GC ISM as a whole)
a multiple scenario is needed to explain the  modest angular resolution
far--IR observations.
The input of chemical models, and higher sensitivity and larger spatial
resolution far--IR observations will lead to a better understanding of the
GC environment.

\acknowledgments

We thank Spanish DGES and PNIE for funding
support under grants PANAYA2000-1784, ESP2001-4516, AYA2002-10113-E,
ESP2002-01627, AYA2003-02785-E and AYA2004-05792. CSO observations were supported by 
NSF grant AST-9980846. 
JRG was also supported by the  french \textit{Direction de la Recherche} in the latest 
stages of the work. We thank J. Le Bourlot for useful
suggestions and for his help with the PDR model.


\clearpage

\begin{deluxetable}{cccccccccc}
\tabletypesize{\scriptsize}
\tablecaption{Summary of H$_2$O spectroscopical and observational data presented in this work.  
Uncertainities refer to fitting errors. The values provided for the 183 GHz line correspond to 
the the direction of Sgr B2(M). In this case, the area of the observed line in K~km~s$^{-1}$ is 
provided instead of the flux. Note that  energy levels
take into account that both \textit{ortho} and \textit{para} ground--states are
at 0 K.
\label{tab-water_lines}}
\tablewidth{0pt}
\tablehead{ 
\colhead{species} & \colhead{transition} & \colhead{$\lambda$($\mu$m)} & \colhead{$\nu$(GHz)} 
& \colhead{$E_{upper}$(K)} &  \colhead{$A_{ij}$(s$^{-1}$)} & F$_{l}$
(W cm$^{-2}$) & F/F$_c$ & $\Delta$v (km s$^{-1}$)  }
\startdata
p--H$_2$O         & $3_{13}-2_{20}$ &1635.439 & 183.3 & 205  & 3.53e$-$06 & 1120$\pm$60 K km s$^{-1}$ & & 30$\pm$3 \\
o--H$_{2}^{18}$O  & $2_{12}-1_{01}$ & 181.053 & 1655.8 &  79 & 5.45e$-$02 & $-$(2.68$\pm$0.26)e$-$18 & 0.73$\pm$0.02 & 52$\pm$2  \\
o--H$_2$O         & $2_{21}-2_{12}$ & 180.488 & 1661.0 & 160 & 2.99e$-$02 & $-$(3.44$\pm$0.46)e$-$18 & 0.74$\pm$0.01 & 64$\pm$3  \\
o--H$_2$O         & $2_{12}-1_{01}$ & 179.527 & 1669.9 &  80 & 5.47e$-$02 & $-$(6.76$\pm$0.29)e$-$14 & 0.06$\pm$0.03 & 204$\pm$23\\
o--H$_2$O         & $3_{03}-2_{12}$ & 174.626 & 1716.7 & 163 & 4.94e$-$02 & $-$(1.86$\pm$0.46)e$-$18 & 0.82$\pm$0.02 & 44$\pm$2  \\
p--H$_2$O         & $3_{13}-2_{02}$ & 138.527 & 2164.1 & 205 & 1.22e$-$01 & $-$(1.27$\pm$0.03)e$-$18 & 0.89$\pm$0.02 & 41$\pm$2  \\
o--H$_2$O         & $4_{32}-4_{23}$ & 121.719 & 2462.9 & 516 & 1.20e$-$01 & $-$(5.40$\pm$0.29)e$-$18 & 0.96$\pm$0.01 & 62$\pm$9  \\ 
o--H$_2$O         & $2_{21}-1_{10}$ & 108.073 & 2773.9 & 160 & 2.52e$-$01 & $-$(6.07$\pm$0.05)e$-$18 & 0.64$\pm$0.02 & 53$\pm$2  \\ 
p--H$_{2}^{18}$O  & $2_{20}-1_{11}$ & 102.008 & 2938.9 & 194 & 2.52e$-$01 & $-$(9.38$\pm$2.62)e$-$19 & 0.96$\pm$0.01 & 72$\pm$8  \\ 
p--H$_2$O         & $2_{20}-1_{11}$ & 100.983 & 2968.7 & 196 & 2.55e$-$01 & $-$(6.02$\pm$0.11)e$-$18 & 0.66$\pm$0.01 & 55$\pm$9  \\
p--H$_2$O         & $3_{22}-2_{11}$ & 89.988  & 3331.4 & 297 & 3.45e$-$01 & $-$(1.99$\pm$0.14)e$-$18 & 0.90$\pm$0.01 & 67$\pm$2  \\ 
o--H$_2$O         & $3_{21}-2_{12}$ & 75.380  & 3977.0 & 271 & 3.25e$-$01 & $-$(4.32$\pm$0.07)e$-$18 & 0.69$\pm$0.02 & 51$\pm$2  \\
p--H$_2$O         & $3_{31}-2_{20}$ & 67.089  & 4468.6 & 410 & 1.20e$+$00 & $-$(1.36$\pm$0.08)e$-$18 & 0.93$\pm$0.03 & 76$\pm$8  \\
o--H$_2$O         & $3_{30}-2_{21}$ & 66.437  & 4512.4 & 376 & 1.22e$+$00 & $-$(8.39$\pm$0.14)e$-$19 & 0.92$\pm$0.02 & 46$\pm$3  \\
\enddata
\end{deluxetable}

\begin{deluxetable}{ccccccccc}
\tabletypesize{\scriptsize}
\tablecaption{Summary of far--IR CO spectroscopical and observational data presented in this work. 
No far--IR CO line has been detected. 3$\sigma$ upper limit line fluxes are tabulated.
\label{tab-co_lines}}
\tablewidth{0pt}
\tablehead{ 
\colhead{species} & \colhead{transition} & \colhead{$\lambda$($\mu$m)} & \colhead{$\nu$(GHz)} 
& \colhead{$E_{upper}$(K)} &  \colhead{$A_{ij}$(s$^{-1}$)} & F$_{l}$ (W cm$^{-2}$) 
& F/F$_c$ }
\startdata
CO & $14-13$ & 185.9  & 1611.8 & 581  & 2.95e$-$04 & $<$2.28e$-$18 & $<$1.007 \\
CO & $15-14$ & 173.6  & 1726.6 & 663  & 3.64e$-$04 & $<$1.02e$-$18 & $<$1.003\\
CO & $16-15$ & 162.8  & 1841.3 & 752  & 4.42e$-$04 & $<$1.14e$-$18 & $<$1.003\\
CO & $17-16$ & 153.3  & 1956.0 & 846  & 5.31e$-$04 & $<$1.49e$-$18 & $<$1.003\\
\enddata
\end{deluxetable}

\end{document}